\documentclass[reprint,aps,pre,showpacs]{revtex4-1}
\usepackage{graphicx}
\usepackage{amssymb}
\usepackage{amsfonts}
\usepackage{amsmath}
\usepackage{mathrsfs}
\usepackage{color}
\usepackage{times}

\newtheorem{theorem}{Theorem}

\newenvironment{proof}[1][Proof]{{\noindent\textbf{#1.}} }{\ \rule{0.5em}{0.5em}}

\newcommand{\qed}{\nobreak \ifvmode \relax \else
      \ifdim\lastskip<1.5em \hskip-\lastskip
      \hskip1.5em plus0em minus0.5em \fi \nobreak
      \vrule height0.75em width0.5em depth0.25em\fi}

\newcommand{\EqLabel}[1]{\label{#1}}
\newcommand{\EqRef}[1]{Eq. (\ref{#1})}
 
\def\ie{{i.e}. }

\newcommand{\lbar}{\left|}

\begin{document}

\title{Logical gaps in the approximate solutions of the social learning game and an exact solution}
\author{Wenjie Dai, Xin Wang, Zengru Di, Jinshan Wu$^{\dag}$ \\School of Systems Science, Beijing Normal University, Beijing 100875, China}

\begin{abstract}
After the social learning models were proposed, finding the solutions of the games becomes a well-defined mathematical question. However, almost all papers on the games and their applications are based on solutions built upon either an add-hoc argument or a twisted Bayesian analysis of the games. Here, we present logical gaps in those solutions and an exact solution of our own. We also introduced a minor extension to the original game such that not only logical difference but also difference in action outcomes among those solutions become visible. 
\end{abstract}

\maketitle

\tableofcontents

\section{Introduction}
The original version of social learning game (see \cite{Bikhchandani2008, Bikhchandani1998} for an introduction and a short review) is a problem with $N$ learners in which each learner (denoted as learner $j$) attempts to identify and act accordingly the true status of a world, which is either in a state $1$ with probability $q^{ext}=0.5$ or in another state $-1$ with probability $1-q^{ext}$, from observing her own private signals ($s^{j}$) and all previous learners' actions ($\vec{a}^{j-1}=\left(a^{1}a^{2}\cdots a^{j-1}\right)$) but without explicitly knowing the previous learners' private signals ($\vec{s}^{j-1}=\left(s^{1}s^{2}\cdots s^{j-1}\right)$). In the game it is assumed that the private signal received by each learner has a probability $p\geq 0.5$ to be the true status of the world. It is usually required that one learner takes an action in every round. Usually, the turn order of learners' action is externally given. A learner receives a positive payoff ($M_{+}$) when her action is the same as the status of the world, and a negative payoff ($-M_{-}$, here $M_{+}=M_{-}$ is assumed, although generally speaking they can be different) otherwise. 

It has been shown\cite{Bikhchandani1992, Banerjee1992} that in this typical setup there is information cascade which can lead to either the proper status of the world or the wrong status even when all the learners are fully rational. After the cascade happens, the rest of all learners choose the same action. Since it is always good to have a better mechanism to encourage cascading towards the true status and reduce the probability of the other, it is necessary to find the equilibriums of this game and to calculate accurately under what conditions such cascading happen.

One key problem in doing so for the social learning game is how, for a learner at the $j$th place to find out the probability distribution of world's status, given the historical record of previous learner's action $\left(a^{1}a^{2}\cdots a^{j-1}\right)$ and her private signal $s^{j}$,
\begin{align}
\lambda^{j} = P\left(s_{w}=1|a^{1}a^{2}\cdots a^{j-1}, s^{j}\right), \EqLabel{eq:lambda}
\end{align}
where $s_{w}$ can be $\pm1$ and it refers to status of the world, $a^{i}=\pm 1$ is the observed $i$th learner's action and $s^{j}$ is the $j$th learner's private signal. Once a learner knows precisely this probability, she can always make an informed decision. Under the assumption that all other learners are as rational and capable as the $j$th learner herself, finding the right formula to calculate $\lambda^{j}$ such that she will get the maximum payoff is a well-defined mathematical problem. 

An exact solution of this mathematical problem refers to a fully rational solution of the above problem from an ideal learner with potentially infinite capability of mathematical calculation. However, except in the case where the private signal is also open to the public, there is not yet an exact procedure to calculate this $\lambda^{j}$. 

A common method of avoiding the calculation of this $\lambda^{j}$ is for the $j$th learner to count the number of actions with an observed value of $1$ (denoted as $N^{j-1,a}_{+}$) and the number of actions with an observed value of $-1$ ($N^{j-1,a}_{-}$) in the previous $\left(j-1\right)$ actions and then to choose to act in concert with the majority after including her own private signals. We call this technique the blind action-counting approach and denote the calculated probability as $\lambda^{j,B}$ in the following. Quite often in theoretical analysis of the social learning game, one focus on the phenomenon of ``cascading'': After observing certain number of previous actions, the rest learners choose the same action no matter what their private signals are. Using the blind action-counting $\lambda^{j,B}$, it is easy to find out that for the original game two consecutive actions, when they are the same, determines action of the next player and thus all the rest players. Therefore, one can study the game by simply enumerating all the cases where cascade happens. However, there is no solid mathematical foundation here to claim that this blind action-counting approach is the best or the exact solution. This approach is commonly used in analysis of the social learning games\cite{Bikhchandani1992, Bikhchandani1998, Callander2009, Guarino2011, Lorenz2007, Neill2005}.

Another commonly used method is based on a Bayesian analysis of the game\cite{Moscarini1998, Smith2000, Bose2006, Bose2008, Gill2008, Liu2012}. We will comment on these two solutions and demonstrate where are the logical gaps in the two solutions in the next section.

The main contribution of this manuscript, besides showing the gaps in the two approximate solutions, is presenting an exact solution of our own. We will first present such a calculation and then compare it against the two approximate solutions on the original game. Although as we will see later our own solution is in principle different from the other two, we will see that there is almost no difference at all among the three solutions on the original social learning game. While we will comment on the reason of this, we propose a minor extension of the social learning game, to which all the solutions should be applicable as well if they are all proper to the original version. We will, however, demonstrate that the three solutions lead to different average payoffs and on average our exact solution has the highest payoff in the extended games. We wish this should be sufficient to illustrate that the two approximate solutions are not as good as the exact ones.

\section{Logical gaps in the blind action-counting and the twisted Bayesian approaches of the social learning game}
\label{sec:logicgap}

In the original definition of the social learning game, all learners know that the true status of the world follows a known priori distribution of all possibilities, which is usually taken as $\left(1, -1\right)$: with a probability $q^{ext} = 0.5$ the status of the world is $1$, \ie,
\begin{align}
q^{ext} = P\left(s_{w}=1\right).
\end{align} 
However, after the world's status is initiated it stays at that status during the entire learning process. After the above-mentioned $\lambda^{j} = P\left(s_{w}=1|a^{1}a^{2}\cdots a^{j-1}, s^{j}\right)$ is known, the rest of decision making process is trivial, \ie $a^{j}=1$ when
\begin{align}
M_{+}\lambda^{j} -M_{-}\left(1-\lambda^{j}\right) > 0 \Longrightarrow \lambda^{j}>\frac{1}{2}.
\EqLabel{eq:sc}
\end{align}
and $a^{j}=-1$ when $\lambda^{j}<\frac{1}{2}$. For the case of $\lambda^{j}=\frac{1}{2}$ additional tie-breaking rules are required, for example
\begin{align}
a^{j} =    \begin{cases}
              1               & \lambda^{j} > \frac{1}{2}\\
              \text{random}(1,-1)              & \text{otherwise}\\
               -1           &  \lambda^{j} < \frac{1}{2}
           \end{cases} 
\EqLabel{eq:randomtiebreak}           
\end{align} 
Here $\text{random}(1,-1)$ means to take one value from $1$ and $-1$ with equal probability. Other tie-breaking rules are also possible\cite{Banerjee1992, Moscarini1998}. 
This relation between $a^{j}$ and $\lambda^{j}$ can also be denoted as 
\begin{align}
a^{j} = \text{sign}\left(\lambda^{j} - \frac{1}{2}\right),
\EqLabel{eq:randomtiebreak2}           
\end{align} 
which in the special case of $\text{sign}\left(0\right)$ is assumed to be $\text{random}(1,-1)$ instead of its usual value $\text{sign}\left(0\right)=0$. In order to use this relation between $a^{j}$ and $\lambda^{j}$ conveniently in later derivations, \EqRef{eq:randomtiebreak} can also be represented by a distribution function of $a^{j}\in\left\{1,-1\right\}$ as
\begin{align}
P\left(a^{j}|\lambda^{j}\right) = \delta\left(a^{j}, \text{sign}\left(\lambda^{j}-\frac{1}{2}\right) \right), 
\EqLabel{eq:delta}
\end{align}
where $\delta\left(i,j\right)$ is the Kronecker $\delta$ notation that it is $1$ when $i=j$ and $0$ otherwise. One can check that \EqRef{eq:randomtiebreak}, \EqRef{eq:randomtiebreak2} and \EqRef{eq:delta} are in fact the same even though the latter takes a form of probability distribution,
\begin{align}
P\left(a^{j}|\lambda^{j}\right) = 
	\begin{cases}
		\begin{cases}
              1              & a^{j}=1\\
                0          &  a^{j}=-1
		\end{cases},
& \lambda^{j}>\frac{1}{2}\\
		\begin{cases}
              \frac{1}{2}              & a^{j}=1\\
                \frac{1}{2}           &  a^{j}=-1
		\end{cases},
& \lambda^{j}=\frac{1}{2}\\
		\begin{cases}
              0             & a^{j}=1\\
              1           &  a^{j}=-1
		\end{cases},
& \lambda^{j}<\frac{1}{2}
	\end{cases}.
\end{align}
The probability distribution form of \EqRef{eq:delta} is very important in deriving our exact procedure. As we will see later, the whole derivation is based on the Bayesian formula, so it is necessary to write all formulae in probability forms.

Now that all of our terminologies and notations have been defined, let us start our discussions on solutions of the social learning game. We have mentioned that the only non-trivial part of the decision making process of the social learning game is the calculation of $\lambda^{j}$. As we stated in the introduction, there are usually two approaches for this calculation. One is to simply count how many times action $1$($-1$) has been taken previously and denoted it as $N^{j-1,a}_{+}$($N^{j-1,a}_{-}$) and compare the two values while taking into account her own private signal $s^{j}$, \ie
\begin{align}
a^{j} = \text{sign}\left(N^{j-1,a}_{+}+s^{j} - N^{j-1,a}_{-}\right).
\EqLabel{eq:Blindshorthand}
\end{align}  
From this formula, it seems that this decision making process does not really need $\lambda^{j}$. We will show later in $\S$ \ref{subsec:Blind} that in fact, it assumes a very special form of $\lambda^{j}$, and under which, \EqRef{eq:Blindshorthand} can be derived. There we will see clearly what is missing in the argument. Here, we first want to present a count argument to point out that there might be better decision making mechanisms other than this blind action-counting approach.
 
Consider the case of a private signal sequence $\left(s^{1}, s^{2}\right) = \left(1, -1\right)$ and the corresponding action sequence being $\left(a^{1}, a^{2}\right) = \left(1, 1\right)$, which is possible under the random tie-breaking rule. Up on observing this action sequence, according to the blind action-counting approach, the third learner will definitely choose the action $a^{3}=1$ no matter what her private signal is. Assuming that her private signal is $s^{3}=-1$, then there is in fact a higher chance that the world is in state $s_{w}=-1$ other than $s_{w}=1$. However, as argued above, the third learner will choose $a^{3}=1$ and thus also the future learners. Therefore, this leads to a wrong cascade. 

Of course, generally speaking when $\left(a^{1}, a^{2}\right) = \left(1, 1\right)$ is observed, it more likely that the world is indeed in a state of $s_{w}=1$, so it is not that wrong to choose action $a^{3}=1$. However, at least, in principle, when the third learner gets $s^{3}=-1$, she should be more careful than simply discarding her own signal especially when she is fully aware of the random tie-breaking rule. Is there any possibility to take this into consideration? The answer is no according to the blind action-counting approximate solution. Will any other solutions be able to take care of this and do better? 

For example, if she is good at mathematics, she should be able to calculate that $P\left(\left(s^{1}, s^{2}\right) = \left(1, 1\right)|\left(a^{1}, a^{2}\right) = \left(1, 1\right)\right)=\frac{2}{3}$ and $P\left(\left(s^{1}, s^{2}\right) = \left(1, -1\right)|\left(a^{1}, a^{2}\right) = \left(1, 1\right)\right)=\frac{1}{3}$ using the Bayesian Formula. Considering her own signal $s^{3}=-1$, she will be less confident that she should choose action $a^{3}=1$ although she might still do. Maybe in some other cases, from the extra bit of inferred information, she will find that the different between the conflicting probability is even smaller than $\frac{2}{3}-\frac{1}{3}$, such that maybe she will choose an action that is different from what she will choose from simply counting $N^{j-1,a}_{\pm}$. As we will see later, this is the whole spirit of this work: Figuring out something like $\vec{s}^{j-1}$ from $\vec{a}^{j-1}$ first and then making better decisions, instead of directly counting $N^{j,a}_{\pm}$ from $\vec{a}^{j-1}$.   

The second commonly used approach\cite{Moscarini1998, Smith2000, Bose2006, Bose2008, Gill2008, Liu2012} of calculation of this $\lambda^{j}$ is more involved than counting $N^{j,a}_{\pm}$. Let us rephrase the formula originally from \cite{Moscarini1998} here in terms of our own notations. Assuming $\xi^{j-1} = P\left(s_{w}=1\lbar\right.\vec{a}^{j-1}\right)$ is known to the $j$th learner for some reasons, which will be explained later, using Bayesian formula, we can have
\begin{widetext}
\begin{align}
\lambda^{j} & = P\left(s_{w}=1|\vec{a}^{j-1}, s^{j}\right) = \frac{P\left(s_{w}=1, s^{j}|\vec{a}^{j-1}\right)}{P\left(s_{j}|\vec{a}^{j-1}\right)}\notag \\
& =  \frac{P\left(s^{j}|s_{w}=1, \vec{a}^{j-1}\right)P\left(s_{w}=1|\vec{a}^{j-1}\right)}{P\left(s_{j}|\vec{a}^{j-1}\right)} \notag \\
& =  \frac{P\left(s_{w}=1|\vec{a}^{j-1}\right)P\left(s^{j}|s_{w}=1\right)}{P\left(s^{j}|s_{w}=1\right)P\left(s_{w}=1|\vec{a}^{j-1}\right) + P\left(s^{j}|s_{w}=-1\right)P\left(s_{w}=-1|\vec{a}^{j-1}\right)}. 
\EqLabel{eq:tB0}
\end{align}
\end{widetext}
This leads to 
\begin{align}
\lambda^{j} = 
\begin{cases}
\frac{\xi^{j-1}p}{\xi^{j-1}p + \left(1-\xi^{j-1}\right)\left(1-p\right)} & s^{j}=1 \\
\frac{\xi^{j-1}\left(1-p\right)}{\xi^{j-1}\left(1-p\right) + \left(1-\xi^{j-1}\right)p} & s^{j}=-1
\end{cases}.
\EqLabel{eq:tB1}
\end{align}
This formula linking $\xi^{j-1}$ to $\lambda^{j}$, while it has a very confusing meaning as we will show latter, is mathematically sound. In order to form a closed formula system, it requires a formula linking $\lambda^{j}$ to $\xi^{j}$, such that the next iteration will give $\lambda^{j+1}$, 
\begin{widetext}
\begin{align}
\xi^{j} \left(s_{w}=1|\vec{a}^{j-1}, a^{j}\right)= \lambda^{j}_{eff}\left(s_{w}=1|\vec{a}^{j-1}, s_{eff}^{j}\right) = 
\begin{cases}
\frac{\xi^{j-1}p}{\xi^{j-1}p + \left(1-\xi^{j-1}\right)\left(1-p\right)} & a^{j}=1 \\
\frac{\xi^{j-1}\left(1-p\right)}{\xi^{j-1}\left(1-p\right) + \left(1-\xi^{j-1}\right)p} & a^{j}=-1
\end{cases}.
\EqLabel{eq:tB2}
\end{align} 
\end{widetext}
In a sense, this assumes that upon observing $a^{j}=1$, the $\left(j+1\right)$th learner will effectively think that $s^{j}=1$ and similarly when observing $a^{j}=-1$. However, this step, exactly this step, is not necessary true.

To summarize, the above Bayesian analysis can be expressed as
\begin{align}
\xi^{j-1} \xrightarrow[s^{j}]{\text{\EqRef{eq:tB1}}}  \lambda^{j} \xrightarrow[]{\text{\EqRef{eq:randomtiebreak}}} a^{j} \xrightarrow[]{\text{\EqRef{eq:tB2}}}  \xi^{j}.
\end{align}
The argument behind \EqRef{eq:tB2} is that when the action $a^{j}=1$ is taken, effectively $s^{j}=1$, no matter what the private signal really is and the same for the case of $a^{j}=-1$. However, while overall, this assumption is not that far off, this is exactly where the logical mistake is. We call the above $\lambda^{j}$ calculated by the above twisted Bayesian approach especially from \EqRef{eq:tB2} the twisted-Bayesian approach and denoted it as $\lambda^{j, tB}$. If we are going to follow this line of thinking we need a better formula from $\lambda^{j}$ to $\xi^{j}$. 

There is another potentially misleading part in the above derivation of \EqRef{eq:tB0}: While it is not mathematically wrong, letting $P\left(s^{j}|s_{w}=1,\vec{a}^{j-1}\right) = P\left(s^{j}|s_{w}=1\right)$ is logically not straight. In the left-hand side, we are thinking that knowing only the action history $\vec{a}^{j-1}$ thus we need to figure out distribution of $s_{w}$ first according to this $\vec{a}^{j-1}$, and then using this `figured out' distribution of $s_{w}$ and limit ourselves in considering only the subset of $s_{w}=1$, and then to calculate probability distribution of $s^{j}$ within the subset of $s_{w}=1$; in right-hand side, we are thinking that when $s_{w}$ is known then the distribution of $s^{j}$ depends only on $s_{w}$ but not on $\vec{a}^{j-1}$. This two expressions are not at the same level of logic. One way of out of this insecure practice of mathematics is to avoid totally $P\left(s^{j}|\vec{a}^{j-1}\right)$ but consider instead things like $P\left(s^{j}|s_{w}=1\right)$ and $P\left(\vec{a}^{j-1}|s_{w}=1\right)$, which are absolutely well-defined. We do so in the next section when constructing exact formula of $\lambda^{j}$.

In one word, the blind action-counting approach does not make use of the full information so that there are rooms for better solutions and the twisted Bayesian analysis missing one important step in its mathematical formalism: There is no solid mathematical ground for \EqRef{eq:tB2} which links $\lambda^{j}$ to $\xi^{j}$.

In the rest of this manuscript, we will present a solution that makes use of the full information and also every step of it has a solid mathematical ground. The only catch is that it is quite mathematically involved and the idea is originated from statistical physics, which might not be a common or familiar toolbox to researchers in social learning, game theory or even other fields of economics. In statistical physics, non-interacting systems are much easier to deal with and quite often it provide a good starting point to build up formalism to tackle interacting systems. It is exactly this beauty of statistical physics that makes it possible to develop our own calculation of $\lambda^{j}$.  

\section{Exact formula of $\lambda^{j}$}

The primary step in a theoretical study of the social learning game is to explicitly solve $\lambda^{j}$ as defined mathematically in \EqRef{eq:lambda} assuming that all learners are fully rational and with infinite capability of mathematical calculation. Using the Bayesian formula, we can rearrange $\lambda^{j}=P\left(s_{w}=1|\vec{a}^{j-1}, s^{j}\right)$ as
\begin{widetext}
\begin{align}
\lambda^{j}  =\frac{P\left(\vec{a}^{j-1}, s^{j}|s_{w}=1\right)P\left(s_{w}=1\right)}{P\left(\vec{a}^{j-1}, s^{j}|s_{w}=1\right)P\left(s_{w}=1\right) + P\left(\vec{a}^{j-1}, s^{j}|s_{w}=-1\right)P\left(s_{w}=-1\right)}  \notag \\ 
=\frac{P\left(\vec{a}^{j-1}|s_{w}=1\right) P\left(s^{j}|s_{w}=1\right)q^{ext}}{P\left(\vec{a}^{j-1}|s_{w}=1\right) P\left(s^{j}|s_{w}=1\right)q^{ext} + P\left(\vec{a}^{j-1}|s_{w}=-1\right) P\left(s^{j}|s_{w}=-1\right)\left(1-q^{ext}\right)}. \EqLabel{eq:Pfinal}
\end{align}
\end{widetext}
Here, in the last step, we have used the fact that the previous actions and the current private signal are two independent events. There is only one unknown term in \EqRef{eq:Pfinal}
\begin{align}
p^{\vec{a}^{j-1}}_{s_{w}}=P\left(\vec{a}^{j-1}|s_{w}\right),
\end{align}
and we denote it as $p^{\vec{a}^{j-1}}_{s_{w}}$, which is the probability of history of a specific previous $j-1$ actions being $\vec{a}^{j-1}$, given world status $s_{w}$. This quantity $p^{\vec{a}^{j-1}}_{s_{w}}$ is different from the $\xi^{j-1} = P\left(s_{w}|\vec{a}^{j-1}\right)$ defined in the twisted Bayesian approach. 

Notice the above $\lambda^{j}$ is subjective, \ie, it is in the $j$th learner's mind that how much she believes the status of the world is $s_{w}=1$ with given information $\vec{a}^{j-1}$ and $s^{j}$. Therefore, $p^{\vec{a}^{j-1}}_{s_{w}}$ is also subjective. In the future in calculating the rate of accuracy and the probability of cascading, we will need an objective probability $\mathcal{P}^{\vec{a}}_{s_{w}}$. The way to find this $\mathcal{P}^{\vec{a}}_{s_{w}}$ is to use
\begin{align}
\mathcal{P}^{\vec{a}}_{s_{w}=1} = \sum_{\vec{s}}p^{\vec{a}}_{\vec{s}}p^{\vec{s}}_{s_{w}=1},
\EqLabel{eq:OaprSw1}
\end{align} 
where $p^{\vec{s}}_{s_{w}=1}$ is totally objective and has nothing to do with learners' decision making and $p^{\vec{a}}_{\vec{s}}$ is the probabilities of all action outcomes $\vec{a}$ given signal sequence $\vec{s}$ and it depends on learners' decision making. The way to calculate $\mathcal{P}^{\vec{a}}_{s_{w}=1}$ is to find out all action outcomes $\vec{a}$ of a given $\vec{s}$ and then sum over all $\vec{s}$ leading to the same $\vec{a}$ according to \EqRef{eq:OaprSw1}. In order to calculate $p^{\vec{a}}_{\vec{s}}$, we need to generate all possible signal sequences $\vec{s}$ and go through the decision-making process, according to given approach of calculating $\lambda^{j}$, to find out action sequences $\vec{a}$ for each of the sequences $\vec{s}$. Notice that $p^{\vec{a}}_{s_{w}=1}$ and $\mathcal{P}^{\vec{a}}_{s_{w}=1}$ are potentially different.

\subsection{A simpler case where private signals are open to the public --- finding $p^{\vec{s}^{j-1}}_{s_{w}}$}
Calculating $p^{\vec{a}^{j-1}}_{s_{w}}$ directly is not easy, however, it is straightforward to calculate $p^{\vec{s}^{j-1}}_{s_{w}}$, 
\begin{align}
p^{\vec{s}^{j-1}}_{s_{w}}=\Pi_{l=1}^{j-1}P\left(s^{l}|s_{w}\right),
\end{align}
where 
\begin{widetext}
\begin{align}
P\left(s^{l}|s_{w}\right)=p^{\frac{1+s^{l}}{2}}\left(1-p\right)^{\frac{1-s^{l}}{2}}\delta_{s_{w},1}
+ \left(1-p\right)^{\frac{1+s^{l}}{2}}p^{\frac{1-s^{l}}{2}}\delta_{s_{w},-1} .
\end{align}
\end{widetext}
Using the signal counting $N^{j-1,s}_{+}$ and $N^{j-1,s}_{-}$, we arrive at
\begin{align}
p^{\vec{s}^{j-1}}_{s_{w}=1}=p^{N^{j-1,s}_{+}}\left(1-p\right)^{N^{j-1,s}_{-}}, \EqLabel{eq:Psignal+} \\
p^{\vec{s}^{j-1}}_{s_{w}=-1}=p^{N^{j-1,s}_{-}}\left(1-p\right)^{N^{j-1,s}_{+}}.\EqLabel{eq:Psignal-}
\end{align}
In terms of these notations, if the private signals are available to the public, then upon receiving a signal $s^{j}=1$ and providing $N^{j-1}_{\pm}$, according to the Bayesian formula,
\begin{widetext}
\begin{align}
\lambda^{j,S} = P\left(s_{w}=1|\vec{s}^{j}\right)
=\frac{P\left(\vec{s}^{j}|s_{w}=1\right)q^{ext}}{P\left(\vec{s}^{j}|s_{w}=1\right)q^{ext} + P\left(\vec{s}^{j}|s_{w}=-1\right)\left(1-q^{ext}\right)},
\end{align}
\end{widetext}
 the probability that the world's status is $s_{w}=1$ can be expressed as
\begin{align}
\lambda^{j, S}\left(\vec{s}^{j-1}, s^{j}\right)
=\frac{1}{1+\left(\frac{1-p}{p}\right)^{\Delta N^{j-1,s}+s^{j}}\frac{1-q^{ext}}{q^{ext}}},
\EqLabel{eq:symmetriclambdasignal}
\end{align}
where $\Delta N^{j-1,s} =\sum_{l=1}^{j-1}s^{l}$. When $q^{ext}=0.5$, this probability is more than $\frac{1}{2}$ as long as $\left(\frac{1-p}{p}\right)^{\Delta N^{j-1,s}+1}<1$. It depends only on $\Delta N^{j-1,s}$. $\lambda^{j}>0.5$ as long as $\Delta N^{j-1,s} + s^{j}>0$. Similar procedure for this public-signal case has also been discussed in \cite{Kleinberg:Networks}.

Next, we generalize the above calculation to the case where only the actions but not the private signals are known to learners.

\subsection{Blind action-counting $\lambda^{j, B}$}
\label{subsec:Blind}
Before that we want to spend a little bit of our time on revisiting of the blind action-counting approach. From the revisiting we will clearly see what information is missing in the blind action-counting approach. 

Even when only previous actions but not private signals are available to the public, let us assume for now that $N^{j-1,a}_{\pm}$includes as much information as $N^{j-1,s}_{\pm}$, thus from \EqRef{eq:symmetriclambdasignal} we have,
\begin{align}
\lambda^{j, B}\left(\vec{a}^{j-1}, s^{j}\right)=\frac{1}{1+\left(\frac{1-p}{p}\right)^{\Delta N^{j-1,a}+s^{j}}\frac{1-q^{ext}}{q^{ext}}},
\EqLabel{eq:Blind}
\end{align}
where $\Delta N^{j-1,a} = \sum_{l=1}^{j-1}a^{l}$. Learners adopting this decision-making mechanism are regarding action sequences provide as much information as signal sequences. This is obviously wrong. It can be shown that according to this formula, $\lambda^{j, B}>\frac{1}{2}$ if $\Delta N^{j-1,a}+s^{j}>0$ and therefore it lead exactly to \EqRef{eq:Blindshorthand}.

In many previous studies of the social learning, calculations of probability of cascades and other quantities were based on this $\lambda^{j, B}$ \cite{Bikhchandani1992, Bikhchandani1998, Callander2009, Guarino2011, Lorenz2007, Neill2005}. This is potentially sub optimal since action sequences are treated as reliable as signal sequences. Next, we are going to present one solution, in which signal sequences and their distributions are figured out first from action sequences and then decision is made upon the inferred signal sequences.

\subsection{From $p^{\vec{s}^{j-1}}_{s_{w}}$ to $p^{\vec{a}^{j-1}}_{s_{w}}$ for exact $\lambda^{j,A}$}

The idea is very simple. If we can turn the history of actions into history of signals, then we can make use of the above $p^{\vec{s}^{j-1}}_{s_{w}}$ and then everything is done. That is to say, we want  all possible signals $\vec{s}^{j-1}$ which lead to action $\vec{a}^{j-1}$. Making use of the law of total probability, we have
\begin{align}
p^{\vec{a}^{j-1}}_{s_{w}} = \sum_{\vec{s}^{j-1}}P\left(\vec{a}^{j-1}|\vec{s}^{j-1}, s_{w}\right)p^{\vec{s}^{j-1}}_{s_{w}}, \EqLabel{eq:Paction}
\end{align}
where $P\left(\vec{a}^{j-1}|\vec{s}^{j-1}, s_{w}\right)$ is related to the decision making process, and which does not directly involves $s_{w}$ because $s_{w}$ is not explicitly known to learners. Therefore,
\begin{align}
P\left(\vec{a}^{j-1}|\vec{s}^{j-1}, s_{w}\right)=P\left(\vec{a}^{j-1}|\vec{s}^{j-1}\right) \equiv p^{\vec{a}^{j-1}}_{\vec{s}^{j-1}}. \EqLabel{eq:signal2action}
\end{align} 
We first notice that because we assume that learners make no mistakes, $\vec{a}^{j-1}$ is fully determined by $\vec{s}^{j-1}$ and the signal vector $\vec{s}^{j-2}$ of previous learners has no direct effect on the current $\left(j-1\right)$th learner's decision. Thus,
\begin{align}
p^{\vec{a}^{j-1}}_{\vec{s}^{j-1}} = P\left(\vec{a}^{j-2}|\vec{s}^{j-1}\right)P\left(a^{j-1}|\vec{s}^{j-1}\right) \notag \\
= P\left(\vec{a}^{j-2}|\vec{s}^{j-2}\right)P\left(a^{j-1}|\vec{s}^{j-2}, s^{j-1}\right)\notag \\
= p^{\vec{a}^{j-2}}_{\vec{s}^{j-2}}P\left(a^{j-1}|\vec{a}^{j-2}, s^{j-1}\right)\notag \\
= p^{\vec{a}^{j-2}}_{\vec{s}^{j-2}}\delta\left(a^{j-1},\text{sign}\left(\lambda^{j-1} \left(\vec{a}^{j-2},s^{j-1}\right)- \frac{1}{2}\right)\right) .  \EqLabel{eq:Gamma}
\end{align} 
In the second last step, we have used the fact that $a^{j-1}$ is not directly determined by $\left(\vec{s}^{j-2}, s^{j-1}\right)$, but is indirectly determined by $\left(\vec{a}^{j-2}, s^{j-1}\right)$, which in turn results from $\left(\vec{s}^{j-2}, s^{j-1}\right)$.

Combining \EqRef{eq:Gamma} and \EqRef{eq:Paction}, we obtain
\begin{widetext}
\begin{align}
p^{\vec{a}^{j-1}}_{s_{w}} = \sum_{\vec{s}^{j-2}, s^{j-1}}p^{\vec{a}^{j-2}}_{\vec{s}^{j-2}}\delta\left(a^{j-1},\text{sign}\left(\lambda^{j-1} \left(\vec{a}^{j-2},s^{j-1}\right)- \frac{1}{2}\right)\right)p^{\vec{s}^{j-2}}_{s_{w}}p^{s^{j-1}}_{s_{w}},  \notag \\
= p^{\vec{a}^{j-2}}_{s_{w}} \sum_{s^{j-1}}\delta\left(a^{j-1},\text{sign}\left(\lambda^{j-1} \left(\vec{a}^{j-2},s^{j-1}\right)- \frac{1}{2}\right)\right)p^{s^{j-1}}_{s_{w}}.
\EqLabel{eq:avecfinal}
\end{align}
\end{widetext}
Here we have used the fact that $p^{\vec{s}^{j-1}}_{s_{w}} = p^{\vec{s}^{j-2}}_{s_{w}}p^{s^{j-1}}_{s_{w}}$. \EqRef{eq:avecfinal} is the central formula of the present work.

\EqRef{eq:Pfinal} and \EqRef{eq:avecfinal} present an iterative procedure to find all $\lambda^{j}$,
\begin{align}
\vec{a}^{j-1} \xrightarrow[\sum_{s^{j-1}}]{\text{\EqRef{eq:avecfinal}}} p^{\vec{a}^{j-1}}_{s_{w}} {\xrightarrow[s^{j}]{\EqRef{eq:Pfinal}} \lambda^{j}\left(\vec{a}^{j-1}, s^{j}\right)\xrightarrow[]{\EqRef{eq:randomtiebreak}} \vec{a}^{j}},
\EqLabel{eq:procedure}
\end{align}
where $\vec{a}^{j-1}$, $p^{\vec{a}^{j-1}}_{s_{w}}$ and the action outcome $a^{j}$ are  public information, while the $\lambda^{j}$ is private to the $j$th learner since $s^{j}$ is hidden from the public. We call the $\lambda^{j}$ calculated from the procedure defined in \EqRef{eq:procedure} the exact solution $\lambda^{j, A}$ since it solve exactly the mathematical problem of finding $P\left(s_{w}|\vec{a}^{j-1}, s^{j}\right)$. Here the Superscript $A$ refers to the fact that it is calculated from histories of actions.

\subsection{Examples showing that $\lambda^{j,A}$ is potentially different from $\lambda^{j,B}$ and $\lambda^{j,tB}$}
\label{subsec:Example}

Next we illustrate this iterative calculation $\lambda^{j,A}$ step by step on the original social learning game. For $\lambda^{1}$, there is no previous $p^{\vec{a}^{0}}_{s_{w}}$ (this quantity may be equivalently set to $1$), so we can directly make use of \EqRef{eq:Pfinal},
\begin{subequations}
\begin{align}
\lambda^{1, A}\left(s_{w}=1|s^{1}=1\right) 
=\frac{1}{1+\left(\frac{1-p}{p}\right)\frac{1-q^{ext}}{q^{ext}}}, \\
\lambda^{1, A}\left(s_{w}=1|s^{1}=-1\right) 
=\frac{1}{1+\left(\frac{1-p}{p}\right)^{-1}\frac{1-q^{ext}}{q^{ext}}}.
\end{align}
\EqLabel{eq:lambda1}
\end{subequations}
It is easy to confirm that at this first step $\lambda^{1, B}$ has exactly the same form as $\lambda^{1, A}$. For  the $\lambda^{j, tB}$ procedure, we start from assuming $\xi^{0} = p\left(s_{w}=1|a^{0}\right) = q^{ext}$, it is easy to confirm from \EqRef{eq:tB1} that
$\lambda^{1, tB} = \lambda^{1, A}$ too.

Following the iterative procedure of $\lambda^{j, A}$, as the second learner we now need to calculate $p^{\vec{a}^{2-1}}_{s_{w}}=p^{a^{1}}_{s_{w}}$ as follows,
\begin{align}
p^{a^{1}}_{s_{w}} =  \sum_{s^{1}}\delta\left(a^{1},\text{sign}\left(\lambda^{1} \left(\vec{a}^{0},s^{1}\right)- \frac{1}{2}\right)\right)p^{s^{1}}_{s_{w}}.
\end{align}
We see that there is a summation over $s^{1}$, because $s^{1}$ is unknown to the second learner. In principle, all values of $s^{1}$ might lead to a given action $a^{1}$ and the corresponding probability is given by $\delta\left(a^{1},\text{sign}\left(\lambda^{1} \left(\vec{a}^{0},s^{1}\right)- \frac{1}{2}\right)\right)$. Considering the first step is just a Kronecker $\delta$ function $\delta_{a^{1}s^{1}}$, \ie, $a^{1}$ has to be exactly $s^{1}$, it is clear that,
\begin{align}
p^{a^{1}}_{s_{w}}  = p^{s^{1}=a^{1}}_{s_{w}}.
\end{align}
With $p^{\vec{a}^{1}}_{s_{w}}$, we are now ready to calculate $\lambda^{2,A}$ using \EqRef{eq:Pfinal}, 
\begin{align}
\lambda^{2,A} \left(s_{w}=1|a^{1}=1,s^{2}=-1\right) 
=\frac{1}{1 + \frac{1-q^{ext}}{q^{ext}}}.
\end{align}
When $q^{ext}=0.5$, this $\lambda^{2,A} \left(s_{w}=1|a^{1}=1,s^{2}=-1\right) =0.5$. Thus an informed decision would be to randomly choose $a^{2}=1$ or $a^{2}=-1$. Similarly we can find all other $\lambda^{2,A}$s. 

It is easy to calculate $\lambda^{2, B}$ from simply counting $\Delta N^{1,a}$ and we find  $\lambda^{2, B}= \lambda^{2, A}$. For $\lambda^{2, tB} \left(s_{w}=1|a^{1}=1,s^{2}=-1\right) $, we need $\xi^{1} \left(s_{w}=1|a^{1}=1\right) $, which equals to $ \lambda^{1,tB}\left(s_{w}=1|s^{2}=1\right)=\frac{1}{1+\left(\frac{1-p}{p}\right)\frac{1-q^{ext}}{q^{ext}}}$. Thus from \EqRef{eq:tB1}, we get
\begin{widetext}
\begin{align}
\lambda^{2,tB} \left(s_{w}=1|a^{1}=1,s^{2}=-1\right)  = \frac{\xi^{1} \left(s_{w}=1|a^{1}=1\right)\left(1-p\right)}{\xi^{1} \left(s_{w}=1|a^{1}=1\right)\left(1-p\right) + \left(1-\xi^{1} \left(s_{w}=1|a^{1}=1\right)\right)p}
=\frac{1}{1 + \frac{1-q^{ext}}{q^{ext}}},
\end{align}
\end{widetext}
which again agrees with $\lambda^{2,A} \left(s_{w}=1|a^{1}=1,s^{2}=-1\right)$. There is no difference among $\lambda^{j,A}$, $\lambda^{j,B}$ and  $\lambda^{j,tB}$ so far.

Proceeding now with the assumption that $a^{2}=1$, we now look at the third learner's decision making process according to either $\lambda^{3,A}$, $\lambda^{3,B}$ or $\lambda^{3,tB}$. Knowing $a^{1}=1, a^{2}=1$, and given, say, $s^{3}=-1$,  $\lambda^{3,B}$ is very straightforward,
\begin{align}
\lambda^{3,B} \left(s_{w}=1|a^{12}=11,s^{3}=-1\right)  = \frac{1}{1 + \left(\frac{1-p}{p}\right)\frac{1-q^{ext}}{q^{ext}}}.
\EqLabel{eq:lambdajB3}
\end{align}
It is more tedious to calculate  $\lambda^{3, tB}$, since it requires  $\xi^{2}$, which according to \EqRef{eq:tB1} in this case of $a^{2}=1$ takes the value of $\lambda^{2,tB} \left(s_{w}=1|a^{1}=1,s^{2}=1\right)= \frac{1}{1 + \left(\frac{1-p}{p}\right)^{2}\frac{1-q^{ext}}{q^{ext}}}$. Then \EqRef{eq:tB1} gives us
\begin{widetext}
\begin{align}
\lambda^{3,tB} \left(s_{w}=1|a^{12}=11,s^{3}=-1\right)  = \frac{1}{1 + \left(\frac{1-p}{p}\right)\frac{1-q^{ext}}{q^{ext}}}=\lambda^{3,B} \left(s_{w}=1|a^{12}=11,s^{3}=-1\right) .
\EqLabel{eq:lambdajtB3}
\end{align}
\end{widetext}
To calculate $\lambda^{3, A}$, we need $p^{\vec{a}^{2}=11}_{s_{w}}$. Considering that $a^{1}=s^{1}$, there are two $\vec{s}^{2}=\left\{11, 1-1\right\}$, both of which possibly lead to $\vec{a}^{2}=11$. Taking into account both possibilities, we obtain
\begin{align}
\lambda^{3,A} \left(s_{w}=1|a^{12}=11,s^{3}=-1\right)  = \frac{1}{1 + \frac{2-p}{1+p}\frac{1-q^{ext}}{q^{ext}}}.
\EqLabel{eq:lambdajA3} 
\end{align}
Notice that this $\lambda^{3,A}$ is different from $\lambda^{3,tB}$, which is the same as $\lambda^{3,B}$. In a sense, the latter considers only the possibility of $\vec{s}^{2}=\left\{11\right\}$.

The difference between \EqRef{eq:lambdajA3} and \EqRef{eq:lambdajB3} may, in principle, lead to different possible actions. When $q^{ext}=0.5$, $p=0.7$, the above $\lambda^{3,A}=0.57$ and $\lambda^{3,B}=0.7=\lambda^{3,tB}$. All the three $\lambda^{3}$s in this case are larger than $0.5$, although their values are different, thus, this difference is not going to generate different actions. However, it is generally speaking possible to derive different actions from  $\lambda^{3,A}=0.57$ and $\lambda^{3,B}=0.7=\lambda^{3,tB}$. Next, we are going to consider such possibilities.

Before further comparing between $\lambda^{j,A}$, $\lambda^{j,B}$ and $\lambda^{j,tB}$, we prove a theorem regarding the conditions of cascading, which later will help us greatly reduce the complexity of calculations.

\subsection{When cascades happen}
\label{sec:cascade}

Cascades have been a central topic in studies of social learning. With the idea behind $\lambda^{j,B}$, being to join the majority, the cascading problem becomes to identify the probability of two consecutive steps with the same actions occurring while the actions before the consecutive two steps are evenly distributed. This is in fact the working criteria of cascading used in many works including the well-known work of \cite{Bikhchandani1992}. In this way, one gets the analytical results of the probability of a cascade before the end of the game $N$,
\begin{subequations}
\begin{align}
P^{+}_{cas} = \frac{p\left(p+1\right)\left[1-\left(p-p^{2}\right)^{\frac{N}{2}}\right]}{2\left(1-p+p^{2}\right)}, \\
P^{-}_{cas} = \frac{\left(p-2\right)\left(p-1\right)\left[1-\left(p-p^{2}\right)^{\frac{N}{2}}\right]}{2\left(1-p+p^{2}\right)}.
\end{align}
\EqLabel{eq:Bikhchandani1992}
\end{subequations}
Instead of $\lambda^{j,B}$, here we want to discuss the phenomenon of cascading based on our exact $\lambda^{j,A}$. We will show that in principle, the fact that two consecutive learners take the same action is not a sufficient condition for cascading if decision is made according to $\lambda^{j,A}$.

\theorem The social learning game starting from the $j$th learner goes into a cascading state $1$ ($-1$) if and only if $\frac{p^{\vec{a}^{j-1}}_{s_{w}=1}}{p^{\vec{a}^{j-1}}_{s_{w}=-1}}>\frac{p}{1-p}\frac{1-q^{ext}}{q^{ext}}$ ($\frac{p^{\vec{a}^{j-1}}_{s_{w}=1}}{p^{\vec{a}^{j-1}}_{s_{w}=-1}}<\frac{1-p}{p}\frac{1-q^{ext}}{q^{ext}}$). 

\proof Cascading happens when $a^{j}=1$ ( $a^{j}=-1$ ) no matter what value the $j$th learner's private signal $s^{j}$ is. First, we find the cascading condition for the $j$th learner. In this proof without loss of generality, we consider only the case of $a^{j}=1$. From \EqRef{eq:Pfinal}, we know  $a^{j}=1$ is equivalent to $\lambda^{j,A}>0.5$. Considering the case $s^{j}=1$ and $s^{j}=-1$, that condition becomes respectively
\begin{align}
\frac{p^{\vec{a}^{j-1}}_{s_{w}=1}}{p^{\vec{a}^{j-1}}_{s_{w}=-1}}>\frac{p}{1-p}\frac{1-q^{ext}}{q^{ext}} \EqLabel{eq:cascade1}\\
\frac{p^{\vec{a}^{j-1}}_{s_{w}=1}}{p^{\vec{a}^{j-1}}_{s_{w}=-1}}>\frac{1-p}{p}\frac{1-q^{ext}}{q^{ext}}  \EqLabel{eq:cascade2}
\end{align}
\EqRef{eq:cascade2} is naturally satisfied if \EqRef{eq:cascade1} is. So the condition holds if and only if  \EqRef{eq:cascade1} is true.
Second,  we want to show that whenever this occurs for the $j$th learner, the same holds for the $\left(j+1\right)$ learner.  From \EqRef{eq:avecfinal}, given that $a^{j}=1$, then
\begin{align}
p^{\vec{a}^{j-1}, a^{j}=1}_{s_{w}} = p^{\vec{a}^{j-1}}_{s_{w}}\left(p^{s^{j-1}=1}_{s_{w}} + p^{s^{j-1}=-1}_{s_{w}}\right)=p^{\vec{a}^{j-1}}_{s_{w}} 
\end{align}
Therefore, \EqRef{eq:cascade1} is again satisfied for $p^{\vec{a}^{j}}_{s_{w}}$. 

\normalfont From this theorem, we know that cascading happens whenever the action taken by a learner does not depend on its private signal. And whenever that happens, later learners can not change this trend of cascading. This simplifies the iterative calculation in that we may stop the iteration when such a cascade is found once, since all the remaining learners will simply choose the same action. This is different from the prior criteria of cascading \cite{Bikhchandani1992} being that two consecutive learners take the same action while the actions before the two were evenly divided between the two actions. In fact, if $\lambda^{j,B}$ is plugged into \EqRef{eq:cascade1}, rather than the exact $\lambda^{j,A}$, we arrive at the prior criteria. However, when $\lambda^{j,A}$ is used, such criteria is no longer sufficient because even after the criteria is met it is possible for the next learner to jump out of the old ``cascade''. In fact, this is intuitively why the rate of accuracy based on $\lambda^{j,A}$ is higher than that of $\lambda^{j,B}$ as we can see later with the RD model from Figure \ref{fig:NscLsa}. 

We identify the place where cascading first happens for each action history $\vec{a}$ and call it as $l_{\vec{a}}$. If no cascading happens for a given process, then we let $l_{\vec{a}}=N$, which is infinity in the large $N$ limit.

\subsection{Comparison between the exact and the two approximate solutions}
\label{subsec:comparison}
With the exact decision making procedure $\lambda^{j, A}$ equipped to every learner, let us now discuss action outcomes of this social learning game. Given values of $p$ and $q^{ext}$, we define as the rate of accuracy as
\begin{align}
r = \sum_{\vec{a}}\mathcal{P}^{\vec{a}}_{s_{w}=1}\left(\frac{1}{N}\sum_{j=1}^{N} \left(a^{j}s_{w}\right)\right), 
\end{align}
where the product $a^{j}s_{w} = 1$ when action $a^{j}$ is the same as $s_{w}$. This $r$ describes average payoff of all learners in a game. 
$\mathcal{P}^{\vec{a}}_{s_{w}=1}$ is the objective probability of the action series $\vec{a}$ as discussed in \EqRef{eq:OaprSw1}. Next we will compare this average payoff of the three decision making procedures according to, respectively $\lambda^{j, A}$, $\lambda^{j, B}$ and $\lambda^{j, tB}$. 

We are also interested in the whole probability of cascading towards respectively $a^{N}=\pm 1$ happening before the last step $N$,
\begin{align}
P^{\left(\pm1\right)}_{cas} = \sum_{\vec{a}, l_{\vec{a}}<N}^{a^{N}=\pm1}\mathcal{P}^{\vec{a}}_{s_{w}=1}.
\EqLabel{eq:Pcas}
\end{align} 

Here, all numerical results will be presented only for the case of $s_{w}=1$ since the cases of $s_{w}=-1$ and $s_{w}=1$ are exactly symmetric. For reasons that will be clear later, all reported results in the following are from games with $N=30$, if not explicitly stated.

\begin{figure}
\includegraphics[width=4cm]{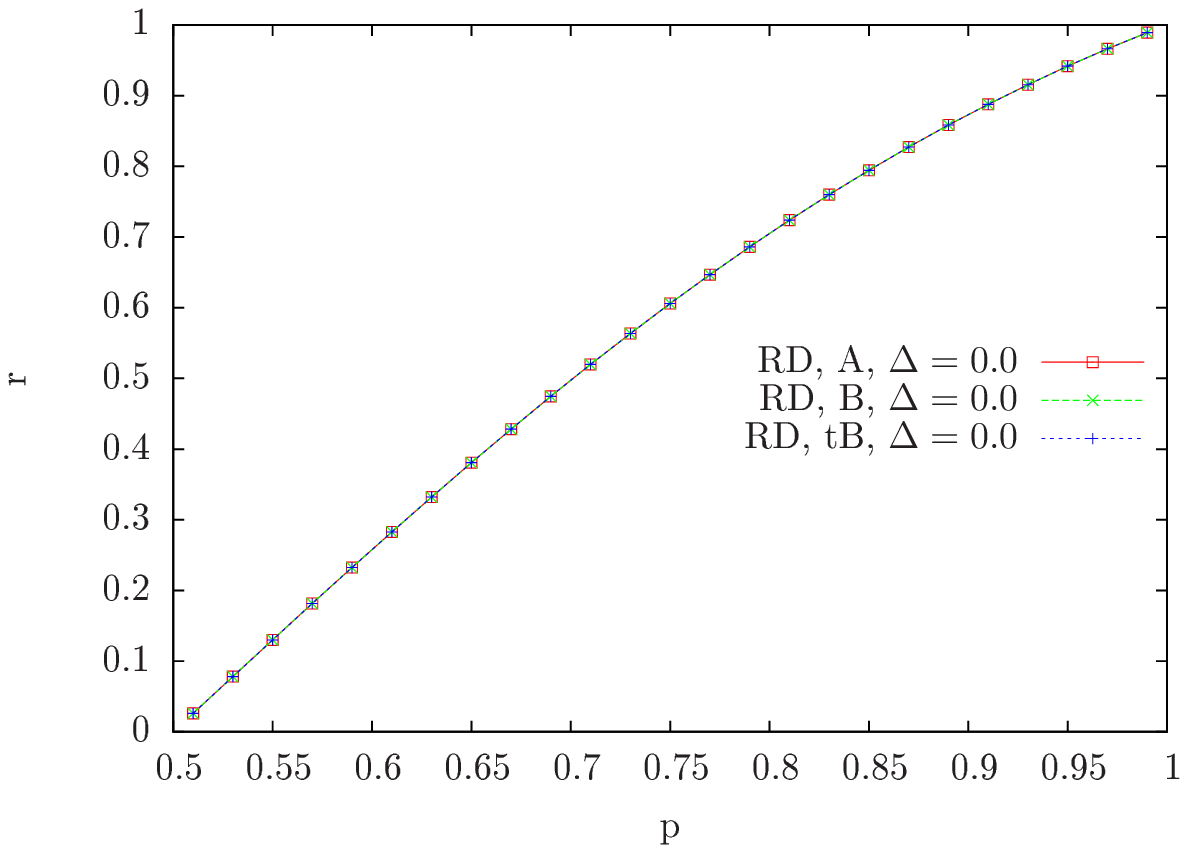} \includegraphics[width=4cm]{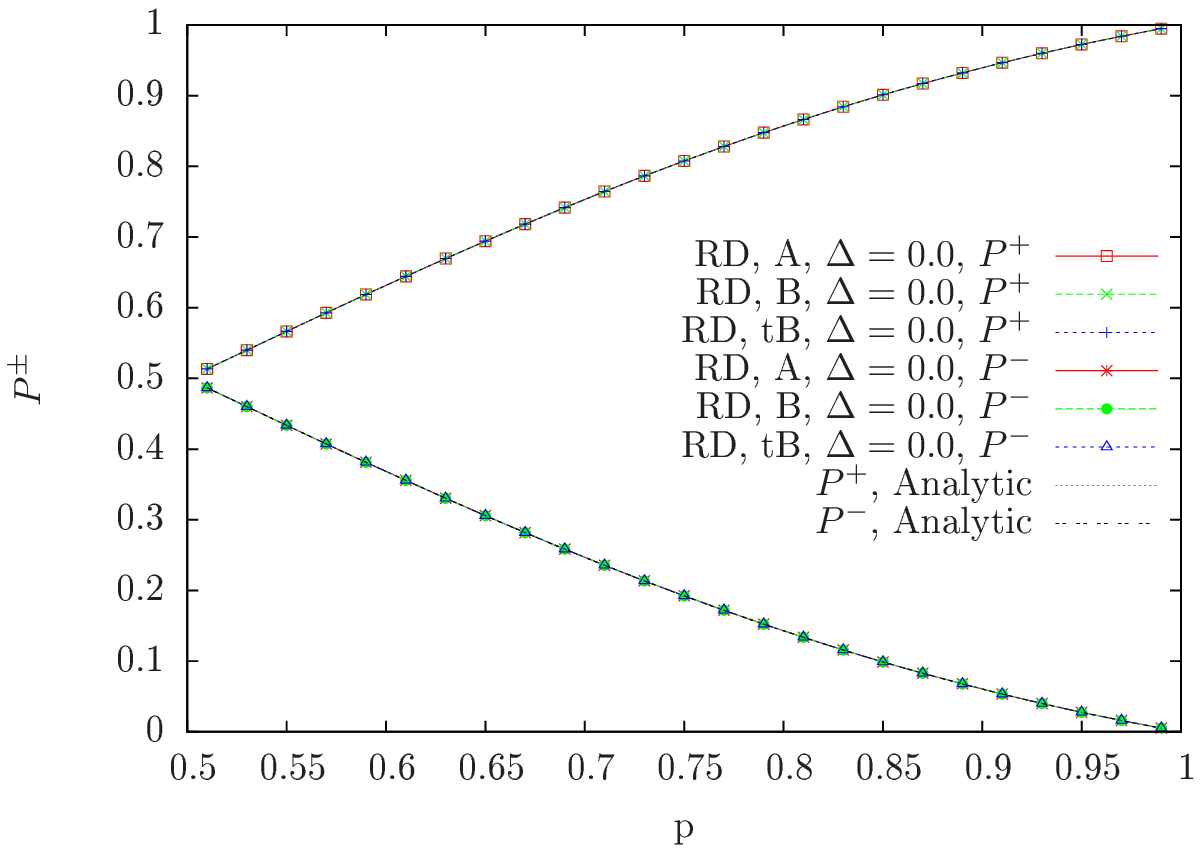} 
\caption{\label{fig:RDlambdas} Rate of accuracy (a) and probability of cascading before $N$ (b) of the original RD models studied using the exact $\lambda^{j, A}$, compared against the blind $\lambda^{j, B}$ and the twisted Bayesian $\lambda^{j, tB}$. In (b), analytical results from \cite{Bikhchandani1992} are also plotted. $N=30$ in this and other figurers, unless noted otherwise.}
\end{figure}

From Fig. \ref{fig:RDlambdas}, we can see that there is no large difference on respectively the rates of accuracy (a) or the probabilities of cascading(a) from all three solutions. In Fig. \ref{fig:RDlambdas}(b), analytical results from \cite{Bikhchandani1992} (also duplicated here as \EqRef{eq:Bikhchandani1992}) are plotted and are shown to exactly agree with our numerical results.

As we can see from \EqRef{eq:lambdajB3}, \EqRef{eq:lambdajtB3} and and \EqRef{eq:lambdajA3}, the three solutions lead to different quantitative values of $\lambda$. However, the qualitative conclusion that $\lambda^{3}$ is larger/smaller than $\frac{1}{2}$ is the same in all three solutions. Therefore, the three quantitatively different expressions give the same action. In this sense, the simple action-counting approach and the twisted Bayesian, while they are not proper in principle, are not bad at all. 

Next, we want to further illustrate that in more general situations different values of \EqRef{eq:lambdajA3},  \EqRef{eq:lambdajB3} and \EqRef{eq:lambdajtB3} may lead to different actions.

\section{Minor modifications of the game and difference between the exact and the approximate solutions}

We have seen that although both $\lambda^{j,B}$ and $\lambda^{j,tB}$ are only approximate solutions while $\lambda^{j,A}$ provide the exact solution of the mathematical problem of finding $\lambda^{j}=P\left(s_{w}|\vec{a}^{j-1},s^{j}\right)$, when applied to real games, there is not much difference among the three solutions. Why? In order to see the difference, situations like a different action is taken after two consecutive same actions should play an essential role in the game playing. In another word, there should be chances for the exact $\lambda^{j,A}$ to correct the wrong cascades. However, for the original social learning game, such situation rarely happens. Here we propose some slight extensions of the game. The extension is so marginal that in principle, if all of them are proper solutions of the original game, the three solutions should also be proper solutions to the extended game. We will, however, show that on the extended game there are visible differences among results of the three solutions.

We introduce a parameter $\Delta$ to represent reservation of a learner to take an action, \ie given $\lambda^{j} = P\left(s_{w}=1|a^{1}a^{2}\cdots a^{j-1}, s^{j}\right)$, a learner $j$ can make a decision according to
\begin{align}
a^{j} = \text{sign}^{\Delta}\left(\lambda^{j}-\frac{1}{2}\right) = 
            \begin{cases}
              1               & \lambda^{j} > \frac{1}{2}+\Delta\\
               \text{random}\left(1,-1\right)               & \text{otherwise}\\
               -1           &  \lambda^{j} < \frac{1}{2}-\Delta
           \end{cases} 
\EqLabel{eq:sign}
\end{align} 
We take $\Delta\in\left[0,\frac{1}{2}\right]$. The same payoff structure applies here, in so far as a learner wins $M_{+}$ when her action is the same as the world's status and loses $M_{-}$ otherwise. Such a reservation can be related with a service charge of taking actions in real game playing. In that case, allowing the learners to take no actions in their turn when $\frac{1}{2}-\Delta < \lambda^{j} < \frac{1}{2}+\Delta$, \ie replacing $\text{random}\left(1,-1\right)$ with simply $0$, makes even a better sense. However, here we will not discuss this additional modification or the motivation of introducing this $\Delta$. What we want to argue is that there is nothing forbids the applicability of each of the three above $\lambda^{j}$s to this extended model if they are applicable to the original one, where simply $\Delta=0$ and it is the only difference between the extended and the original games. From where we stand, we simply want the difference between \EqRef{eq:lambdajA3},  \EqRef{eq:lambdajB3} and \EqRef{eq:lambdajtB3} to make a difference on action outcomes.

Next, we want to compare results of the three solutions on the extended model with $\Delta\neq 0$. 

\subsection{$\lambda^{j,A}$ is different from $\lambda^{j,B}$ and $\lambda^{j,tB}$ when $\Delta>0$}

In $\S$\ref{subsec:Example} and $\S$\ref{subsec:comparison} we have shown that although the resulted formulae and the numerical values are different among $\lambda^{j,A}$, $\lambda^{j,B}$ and $\lambda^{j,tB}$, the resulted actions are not different at all when $\Delta=0$. Now let us do the same comparison when $\Delta\neq 0$. According to \EqRef{eq:lambdajA3},  \EqRef{eq:lambdajB3} and \EqRef{eq:lambdajtB3}, when $\Delta=0.1$ these two generate different actions: The third learner choose to act randomly according to $\lambda^{j,A}$ and choose action $1$ according to $\lambda^{j,B}$ and  $\lambda^{j,tB}$. 

\begin{figure}
\includegraphics[width=4cm]{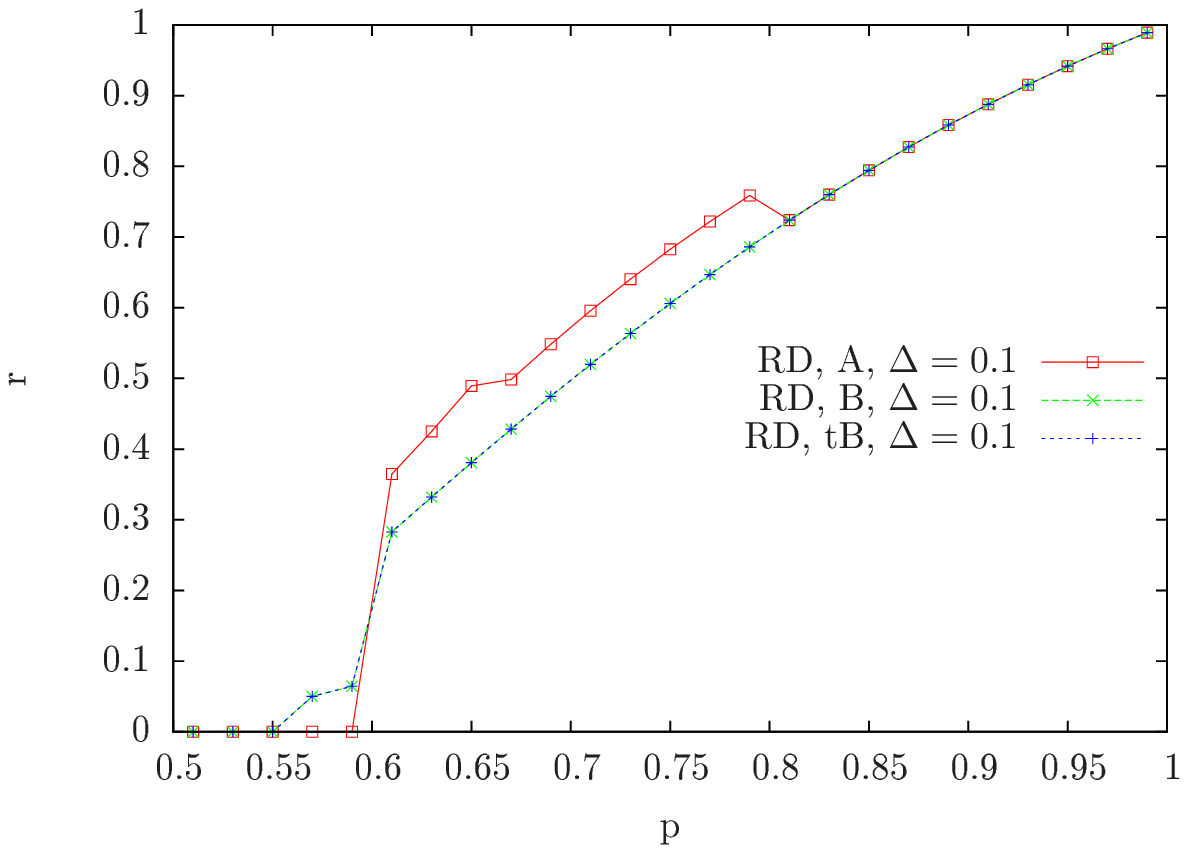}\includegraphics[width=4cm]{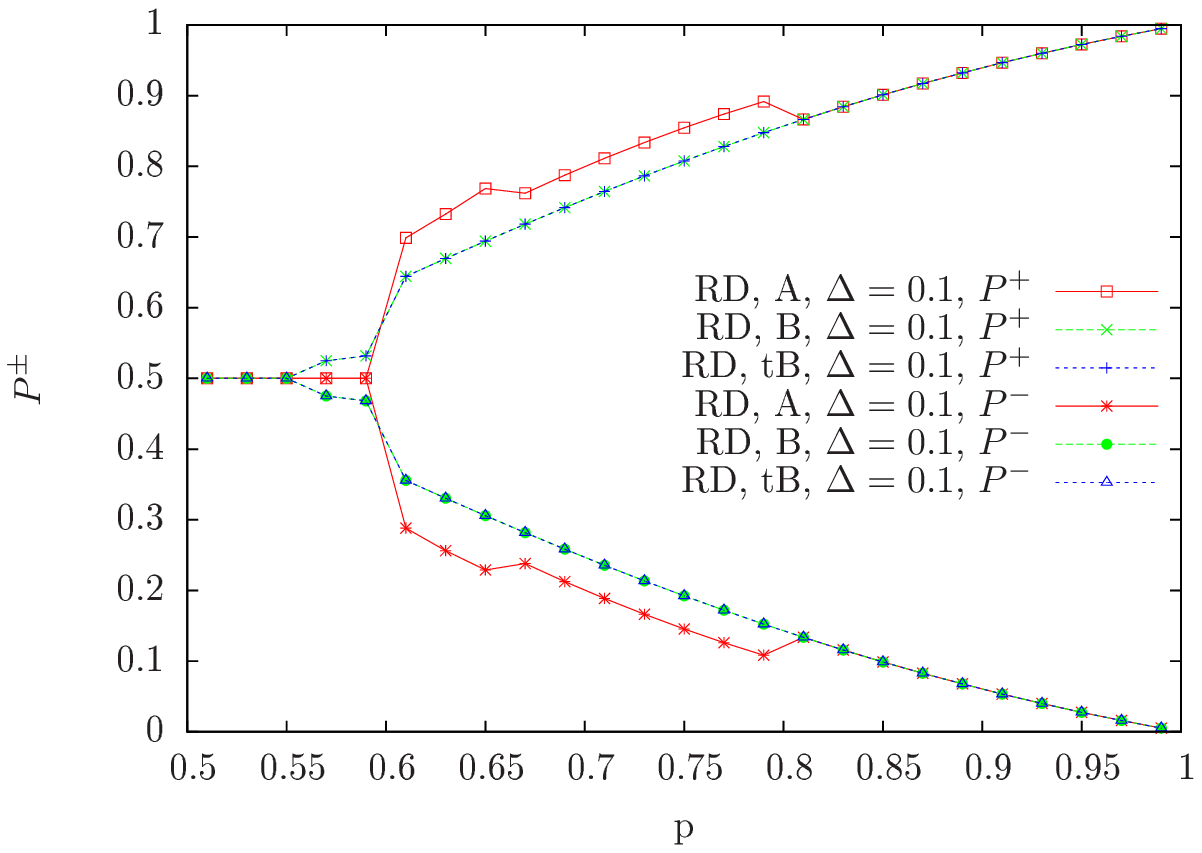} 
\caption{\label{fig:RDlambdasDelta} Rate of accuracy (a) and probability of cascading before $N$ (b) of the extended model with $\Delta=0.1$ solved by the exact $\lambda^{j, A}$, the blind $\lambda^{j, B}$ and the twisted Bayesian $\lambda^{j, tB}$.}
\end{figure}

We also see manifestation of this difference in simulation results. Where $\Delta=0.1$, we see that rate of accuracy and the probability of cascading calculated from $\lambda^{j,A}$, shown in Fig. \ref{fig:RDlambdasDelta}, are different from and in fact higher than those calculated from  $\lambda^{j,B}$ and $\lambda^{j,tB}$. Therefore, action sequences from the strategic state $\lambda^{j,A}$ must be different from those from $\lambda^{j,B}$ or $\lambda^{j,tB}$.

We have shown in Table \ref{tab:actions} a list of possible action outcomes according to all three procedures of $\lambda^{j}$ for a specific signal sequence, where we can clearly see that when $q^{ext}=0.5$, $\Delta=0.1$, $p=0.7$ the action outcome from $\lambda^{j,A}$ is different from those from $\lambda^{j,B}$ and $\lambda^{j,tB}$. For this given $\vec{s}$, it is more likely that $s_{w}=-1$. Here we assume that the second learner, upon calculating $\lambda^{2}=0.5$, randomly chose $a^{2}=1$. This leads to a wrong cascades in the cases of $\lambda^{j,B}$ and $\lambda^{j,tB}$. However, when $\lambda^{j,A}$ is used, there is a chance that such a false cascade can be corrected depending on whether the third learner chose $a^{3}=1$ or $a^{3}=-1$ when $\lambda^{3,A}\left(s_{w}|\left(a^{1},a^{2}\right)=\left(1,1\right), s^{3}=-1\right)-\frac{1}{2}<\Delta$.
\begin{center}
\begin{table}
	\caption {Possible action outcomes following different $\lambda^{j}$s. $q^{ext}=0.5$, $\Delta=0.1$, $p=0.7$. } \label{tab:actions} 
	\begin{tabular}{ {r}| l*{3} {c}}
	\hline
    	$\vec{s}$ & 1 &  -1 & -1 & -1 \\ \hline
    	$\vec{a},\lambda^{j,A}$ & 1 &1  &-1  &-1   \\   \hline
    	$\vec{a},\lambda^{j,B}$ & 1 &1  &1  &1   \\  \hline
    	$\vec{a},\lambda^{j,tB}$ &1 &1  &1  & 1   \\
	\hline
  	\end{tabular}
\end{table}
\end{center}

\subsection{Emergence of pseudo-cascade and number of possible signals corresponding to a given action outcome}

It has been conjectured that cascade happens when two consecutive learners take the same action while the number of $a=1$ and $a=-1$ are the same in actions of previous learners, $a^{j-1}=a^{j}$ while $N^{j-2,a}_{+}=N^{j-2,a}_{-}$. We call this a pseudo-cascade. This conjecture will be true if $\lambda^{j,B}$ is an exact solution since when two consecutive actions are the same, the next learner's private signal will not be able to change the sign of $\Delta N^{a}$. We have proved earlier that a real cascade, which is a trap state that no later learners can ever come out of, should be defined a learner choosing the same action no matter what the private signal a learner receives is. These two definitions are potentially different. In this section, we want to investigate the difference between the two. We will record the number of pseudo-cascade emerging and count it towards $n_{sc}\left(\vec{a}\right)$ if the action after that consecutive same action sequence is different. We want to know on average how often this happens, 
\begin{align}
N_{sc} = \sum_{\vec{a}} n_{sc}\left(\vec{a}\right)\mathcal{P}^{\vec{a}}_{s_{w}}.
\end{align}
If the conjecture about pseudo-cascade is right then $N_{sc}=0$.      

\begin{figure}
\includegraphics[width=5cm]{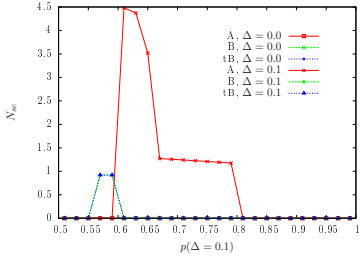}
\caption{\label{fig:NscLsa} Average number of pseudo cascading happening before real cascading. $q^{ext}=0.5$.}
\end{figure}

From Fig. \ref{fig:NscLsa}, we find that when $\Delta=0$, $N_{sc}=0$ and there is no difference between the three decision-making procedures. This implies that in this case, all pseudo-cascades are in fact real cascades. When $\Delta=0.1$, $N_{sc}>0$ appears in action outcomes from all three solutions and the one corresponding to $\lambda^{j,A}$ has the largest value. This means that there are more pseudo-cascades corrected by the $\lambda^{j,A}$ than the other two. This can be seen as the reason why the average payoff from $\lambda^{j,A}$ is higher than the other two.

\subsection{On average, when cascades happen}
\begin{figure}
\includegraphics[width=4cm]{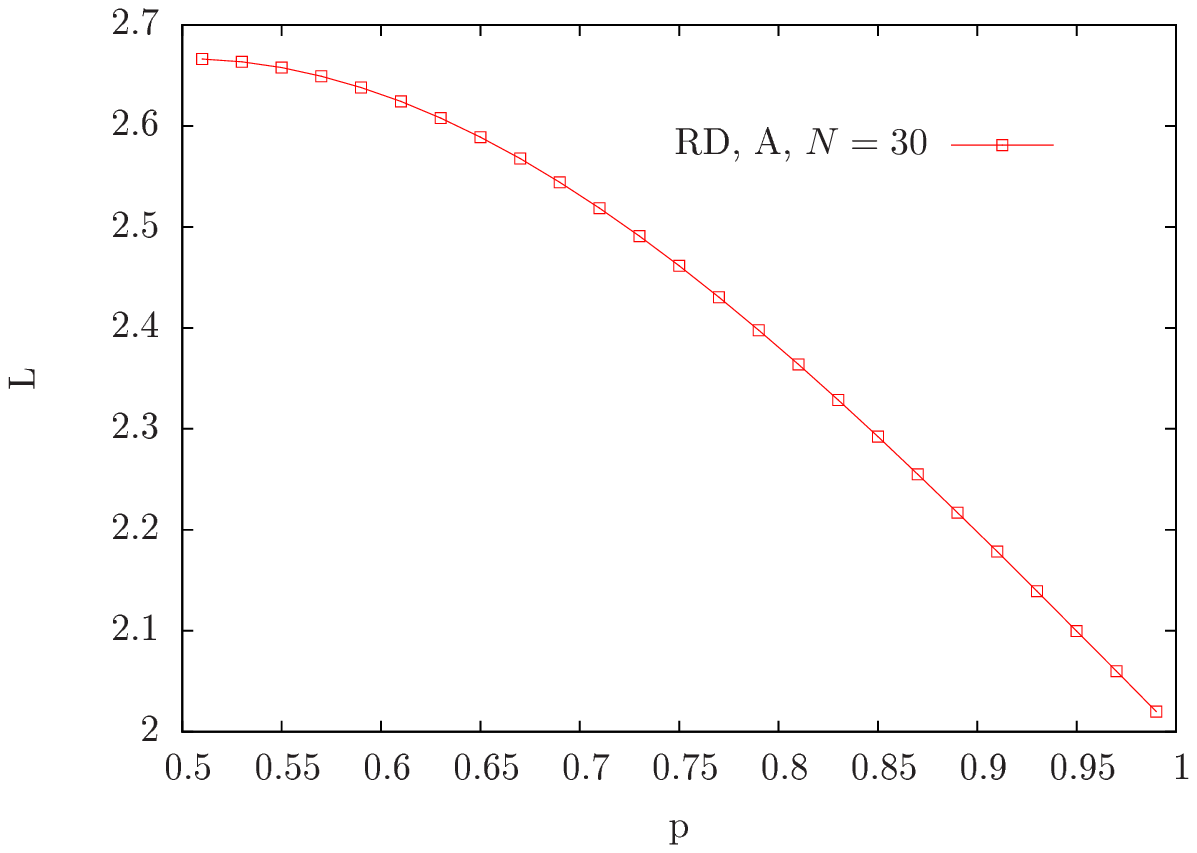}\includegraphics[width=4cm]{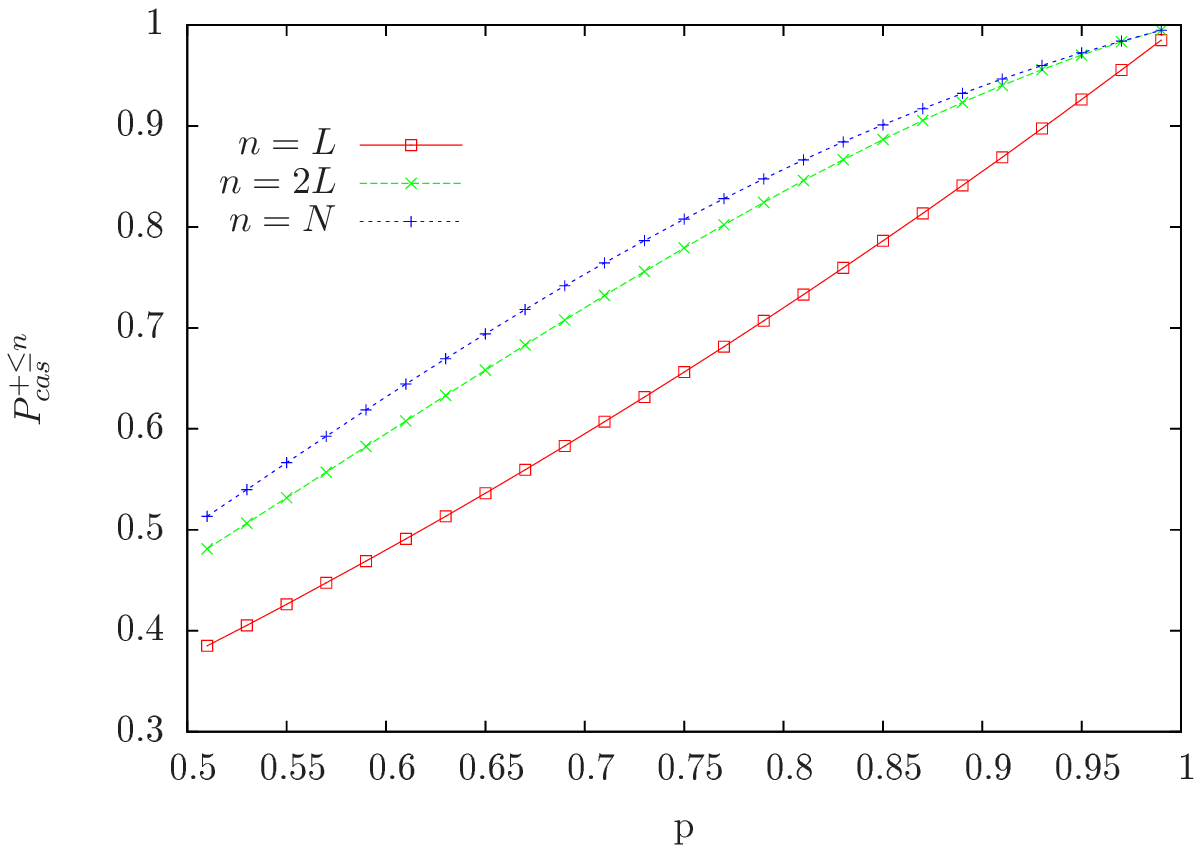} 
\caption{\label{fig:RDL}Average cascading length (a) and probability of cascading before the $n$th step (b) solved by the exact solution when $\Delta=0, q^{ext}=0.5$. }
\end{figure}

In our simulation, we used a relatively small size of population, $N=30$. In this section, we want to present evidence that there is no need to go much further beyond that. Another interesting question is when is a good time for learners to take actions if they have certain degree of freedom in choosing when to act. One might intuitively guess that generally it is better to act later than earlier and even if there is a cost of waiting one should look to determine the best timing to act by comparing the cost and the gain \cite{Gale1994_timing}. Here we try to provide another perspective. We define an average cascading length,
\begin{align}
L = \sum_{\vec{a}} l_{\vec{a}} \mathcal{P}^{\vec{a}}_{s_{w}=1},
\end{align} 
and a probability of cascading before the $n$th step,
\begin{align}
P^{\left(\pm1, 0\right), \leq n}_{cas} = \sum_{\vec{a}, l_{\vec{a}}\leq n}^{a^{n}=\pm1, 0}\mathcal{P}^{\vec{a}}_{s_{w}=1}.
\EqLabel{eq:Pcasn}
\end{align} 
Here $l_{\vec{a}}$ is defined in $\S$\ref{sec:cascade} and it denotes the position of the first learner whose private signal does not make any difference in action. \EqRef{eq:Pcasn} is in fact a generalization of \EqRef{eq:Pcas}, where $n$ takes a specific value $N$. We plot $L$ vs. $p$ and $P^{+,\leq n}_{cas}$ vs. $p$ in Fig. \ref{fig:RDL}. We see from Fig. \ref{fig:RDL}(a) that for all values of $p$, $L$ is always much smaller than $N$, which means that on average cascading happens far before the end of the game. We also find from Fig. \ref{fig:RDL}(b) that $P^{+,\leq2L}_{cas}$ is very close to $P^{+,\leq N}_{cas}$ although $P^{+,\leq L}_{cas}$, the cascading probability before the average length $L$ is considerably smaller than the whole cascading probability $P^{+,\leq N}_{cas}$. This means that choosing to act near $2L$ is on average long enough that whatever cascading is going to happen it likely happened already. Notice that $2L$ is still much smaller than $N$. This defined average cascading length is also meaningful in determining scales of advertising/manipulation. For example, in an experiment of two economists secretly buying one of their own books, determining the minimum but sufficient number of copies initially purchased is an important task. We believe this $L$ is a good index for this purpose.

\begin{figure}
\includegraphics[width=5cm]{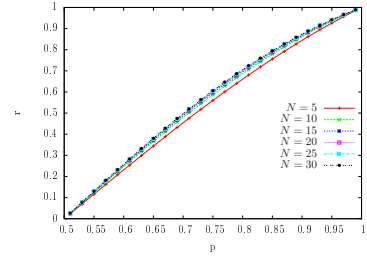}
\caption{\label{fig:rvsN}Rate of accuracy found from the exact solution for various values of $N$. $\Delta=0, q^{ext}=0.5$.}
\end{figure}

Note that $L\ll N$ and $P^{+, \leq 2L}_{cas}$ is already very close to $P^{+,\leq N}_{cas}$. This implies that for our numerical study to obtain an accurate overall picture of the game, ideally with infinite $N$, it is in fact not necessary to have $N$ be too much larger than several times $L$, which for the maximum case is around $3$. To test this further, we plot the rate of accuracy found from the exact solution for various values of $N$ in Fig. \ref{fig:rvsN}. This plot shows that the difference between $N=25$ and $N=30$ is very small. This confirms that it was reasonable to set $N=30$ for all previous game results. 

\section{Conclusions and discussion}

An exact solution to the original social learning game is proposed and logical gaps in the two approximate but commonly used solutions are discussed. In order to demonstrate the three solutions are not only different in principle but also lead to different outcomes in actions and payoffs, we modified the game to incorporate a parameter $\Delta$, which stands for level of reservation for risk taking. 

We have also confirmed via numerical results from simulations that action outcomes and thus the received payoffs from $\lambda^{j,A}$ is different from the outcomes from $\lambda^{j,B}$ or $\lambda^{j,tB}$.

Why the two approximate solutions work so well on the original game? We have shown that $\lambda^{j,B}$ is equivalent to assume that the game enters into a cascade after two consecutive learners take the same action and we have also discussed that the difference between $\lambda^{j,A}$ and $\lambda^{j,B}$ is that there is a chance of getting out of wrong cascades when $\lambda^{j,A}$ is used. This is in fact why $\lambda^{j,A}$ is potentially better than $\lambda^{j,B}$, where there is no such a chance of correction. However, we have seen that there is not a big chance of happening of such events of getting out of cascading in real game playing when $\Delta=0$. This is why $\lambda^{j,B}$, while not exact, is not that bad at all. When it comes to $\lambda^{j,tB}$, the assumption that $s^{j}_{eff}=a^{j}$,  while it is not exact either, is not that far off. Upon observing $a^{j}=1$ it is a safe guess to take $s^{j}=1$ too and this -- regarding $s^{j}$ as $a^{j}$ -- makes $\lambda^{j,tB}$ equivalently $\lambda^{j,B}$. That is why these inexact solutions are more or less acceptable. However, we want to emphasize again that in principle, there are logical gaps in both.

It is very costly, in the sense that large amount of tricky mathematical calculation is required to adopt $\lambda^{j,A}$, for a learner to really implement the exact solution. How close are results from our human decision making processes to the exact solution? This is another interesting question. A trivial solution to this is that once learners understand that this more involved formula is what is needed to play the social learning game, they can always use computer to help them in decision making. A highly non-trivial question will be, in game playing processes in the real world, should we expect that action outcomes to be close to those predicted by $\lambda^{j,A}$, or the other two? This should be a question of further investigation.

\bibliographystyle{naturemag}
\bibliography{SocialLearning}
\end{document}